\pdfoutput=1
\documentclass{article} 
\usepackage{iclr2022_conference,times}

\usepackage{amsmath,amsfonts,bm}

\def\eqref#1{equation~\ref{#1}}

\def\1{\bm{1}}

\def\rvh{{\mathbf{h}}}

\def\rvz{{\mathbf{z}}}

\def\va{{\bm{a}}}

\def\ve{{\bm{e}}}

\def\vh{{\bm{h}}}

\def\vn{{\bm{n}}}

\def\vv{{\bm{v}}}

\def\vx{{\bm{x}}}

\def\vz{{\bm{z}}}

\DeclareMathAlphabet{\mathsfit}{\encodingdefault}{\sfdefault}{m}{sl}
\SetMathAlphabet{\mathsfit}{bold}{\encodingdefault}{\sfdefault}{bx}{n}

\def\gE{{\mathcal{E}}}

\def\gS{{\mathcal{S}}}
\def\gT{{\mathcal{T}}}

\def\gX{{\mathcal{X}}}

\def\sR{{\mathbb{R}}}
\def\sS{{\mathbb{S}}}

\DeclareMathOperator*{\argmin}{arg\,min}

\usepackage{hyperref}
\usepackage{url}
\usepackage{graphicx}
\usepackage{makecell}
\usepackage{multirow}
\usepackage{enumitem}
\usepackage{bbm}
\usepackage{color,xcolor}
\usepackage{booktabs}
\usepackage{lineno}
\usepackage{array}
\usepackage{ulem}
\usepackage{multicol}
\usepackage{amsmath,amsthm,amsfonts,amssymb,bm}
\usepackage{bbding}
\usepackage{stfloats}
\usepackage[font=small,labelfont=bf]{caption}
\usepackage{tabulary}
\usepackage{tabularx}
\usepackage{float}

\title{VQPL: Vector Quantized Protein Language}

\author{Zhangyang Gao$^{\dagger}$, Cheng Tan$^{\dagger}$, Stan Z. Li$^{*}$\\
AI Lab, Research Center for Industries of the Future, Westlake University \\
\texttt{\{gaozhangyang, tancheng,Stan.ZQ.Li\}@westlake.edu.cn}\\
\thanks{Preprint version. $^{\dagger}$Equal Contribution, $^{*}$Corresponding Author.}
}

\makeatletter
\def\thanks#1{\protected@xdef\@thanks{\@thanks
        \protect\footnotetext{#1}}}
\makeatother

\iclrfinalcopy 
\begin{document}
\maketitle

\begin{abstract}
   Is there a foreign language describing protein sequences and structures simultaneously? Protein structures, represented by continuous 3D points, have long posed a challenge due to the contrasting modeling paradigms of discrete sequences. To represent protein sequence-structure as discrete symbols, we propose a VQProteinformer to project residue types and structures into a discrete space, supervised by a reconstruction loss to ensure information preservation. The sequential latent codes of residues introduce a new quantized protein language, transforming the protein sequence-structure into a unified modality. We demonstrate the potential of the created protein language on predictive and generative tasks, which may not only advance protein research but also establish a connection between the protein-related and NLP-related fields. The proposed method will be continually improved to unify more protein modalities, including text and point cloud. 
\end{abstract}

\section{Introduction}
Recent years have witnessed remarkable progress in learning protein sequence and structure representations in the fields of protein mutation prediction \citep{umerenkov2022prostata}, protein sequence design \citep{ingraham2019generative, jing2020learning, tan2022generative, gao2022alphadesign, hsu2022learning, gao2023pifold}, and protein property prediction \citep{zhang2022protein,fan2022continuous}. However, the modality gap between sequence and structure usually requires different model architectures. For example, sequence learners are typically based on transformers, inspired by the great success of natural language processing (NLP). On the other hand, the structure learners are generally based on graph neural networks (GNNs) to consider the spatial interactions between residues. As a result, the modality gap makes it inefficient and complex to learn the joint representation of sequence and structure. In addition, the GNN structure learners limit the capability of extending to large-scale, due to the problems of oversmoothing, computing overhead and the sub-optimal expressive power compared to transformers. \textit{Is there a way to unify the sequence and structure modalities?}

Recent work shows promise for unifying protein modalities. For example, FoldingDiff \citep{wu2022protein} suggests converting protein structures as sequences of folding angles and shows great potential in unconditional protein structure generation. Further, DiffSDS \citep{gao2023diffsds} extends the method in the scenarios of conditional and constrained protein structure generation. The angle sequences have greatly simplified the model design, where advanced transformers can be directly applied to learn the protein structure space. Surprisingly, the experimental results show that the angle sequences can achieve comparable performance with the 3D coordinates. Although these works demonstrate the effectiveness of learning structures via angle sequences, the continual angle values cannot be stored as discrete symbols like human language. Learning a discrete protein language for connecting protein-related research with NLP-related work and improving tasks such as multi-modal protein design and protein property prediction remains an open problem.

Inspired by the success of learning discrete vision language \citep{yu2021vector, esser2021taming, }, we propose a novel protein language that represents sequence and structure simultaneously. The key idea is to project the protein sequence and structure into a discrete space via vector quantized transformers (VQTs), then the discrete latent codes can serve as a new protein language to unify the sequence and structure modalities. We conducted preliminary evaluations of the created protein language in both predictive and generative tasks.

\section{Method}

\subsection{Overall framework}
\begin{figure}[h]
   \centering
   \includegraphics[width=5.5in]{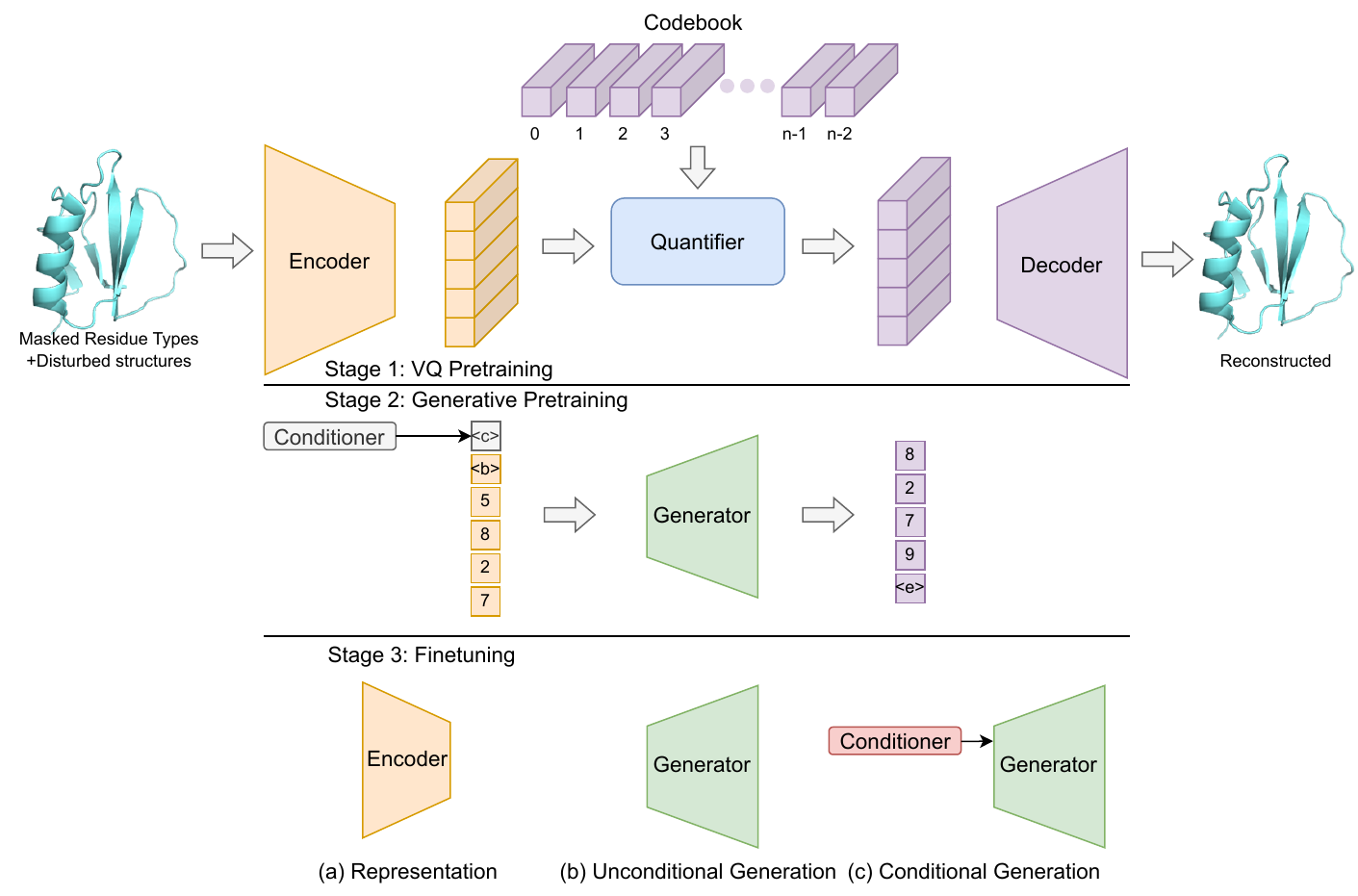}
   \caption{The overall framework of MPL. }
   \label{fig:framework}
\end{figure}

As shown in Figure.\ref{fig:framework}, we introduce a VQProteinformer (encoder-quantizer-decoder model) to jointly encode the residue types and structures into a discrete latent space and reconstruct them from the discrete latent codes. Given the protein sequence and structure as $\gS = \{s_i: 1 \leq i \leq n\}$ and $\gX = \{\vx_i \in \sR^{3}: 1 \leq i \leq n\}$, where $n$ is the protein length, the VQProteinformer could be formulated as:

\begin{equation}
   \label{eq:VQProteinformer}
   \begin{cases}
      \rvh \sim  q(\rvh| \gT(\gS), \gT(\gX))\\
      \rvz = \text{quantizer}(\rvh)\\
      \hat{\gS}, \hat{\gX} = p(\gS, \gX|\rvz)\\
   \end{cases}
\end{equation}

The encoder $q(\cdot)$ is a transformer-based model that takes the protein sequence and structure as input and outputs the continuous latent embedding $\rvh = \{\vh_1, \vh_2, \cdots, \vh_n\} \in \sR^{n,d}$. The quantizer is a vector quantizer (VQ) module that maps the continuous embeddings $\rvh$ into discrete latent code $\rvz = \{z_1, z_2, \cdots, z_n\} \in \sR^{n}$. The decoding transformer $p(\cdot)$ recovers the sequence and structure from the discrete latent codes $\rvz$. The $\gT(\cdot)$ indicates data corruption operations such as masking and noising on the sequence and structure to force the model learn non-trivial representations.

\subsection{Learning Protein Language}
\begin{figure}[h]
   \centering
   \includegraphics[width=3.5in]{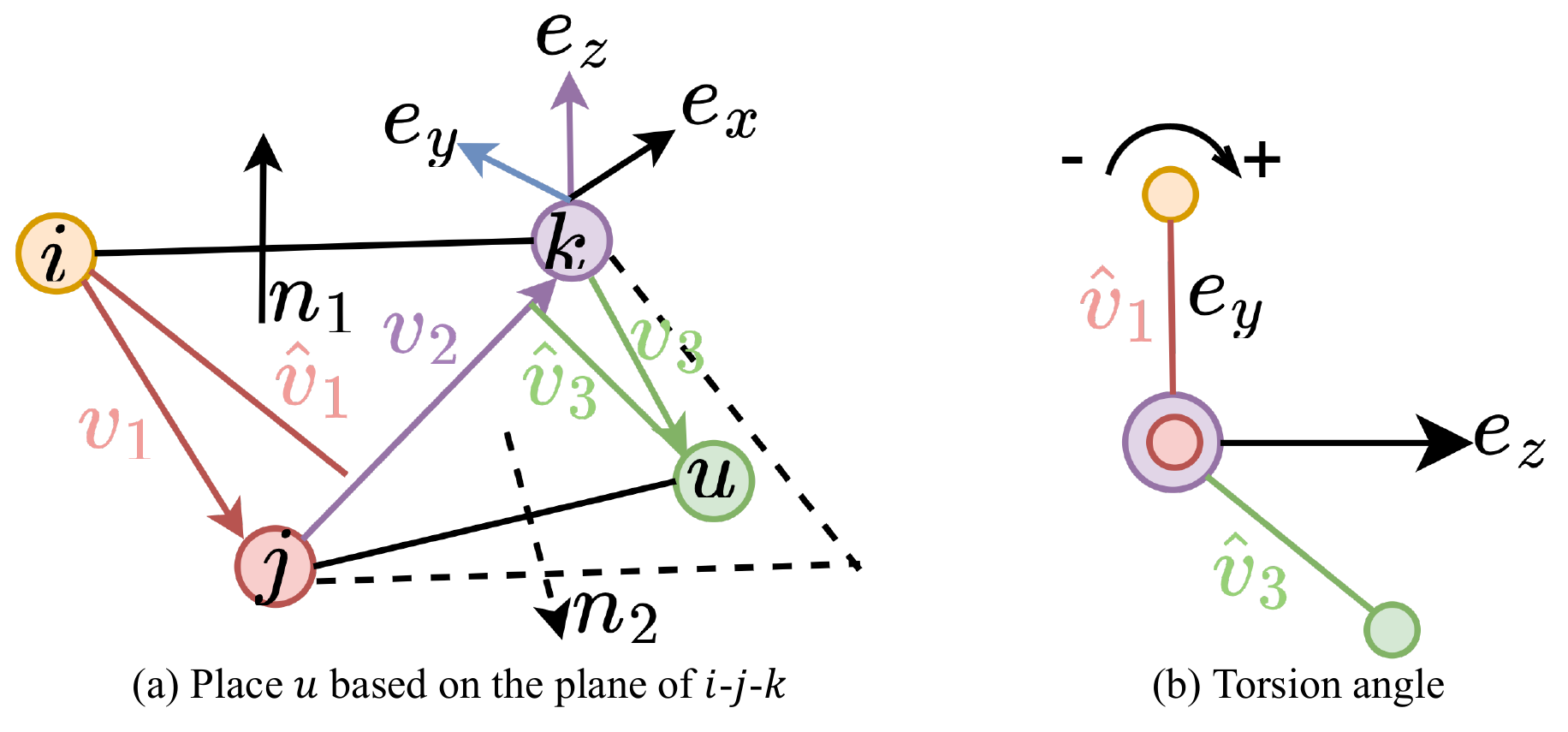}
   \caption{Torsion angle computation. }
   \label{fig:torsion_angle}
\end{figure}

\paragraph{Sequential Protein Representation} The natural language is a left-to-right sequence, which is also suitable for protein sequences. However, protein structures are typically represented as graphs with equivariant node features \citep{gao2022alphadesign}. This introduces additional degrees of freedom and necessitates different modeling paradigms compared to sequence modeling. To address this problem, we convert the 3d points into a sequence of angles, which is rotation and translation invariant, for describing the intrinsic protein structure. As shown in Fig.~\ref{fig:torsion_angle}, following the backbone direction, given the previous three atoms, we can construct a local frame $T_{u} = \{\vx_i, \vx_j, \vx_k\}$. The next atom $\vx_u$ can be represented by the distance $r_u$ and two torsion angles $\alpha_u$ and $\beta_u$, denoted as $\va_u = (r_u, \alpha_u, \beta_u)$. We provide the computational details in the appendix.

\paragraph{Protein Encoder} The encoder is a transformer-based model that takes the corrupted protein sequence $\gT(\gS)$ and structure $\gT(\gX)$ as input and outputs the continuous latent codes $\rvh$. The corruption operations $\gT(\cdot)$ include masking and noising on the sequence and structure, which will be described in the experimental sections. The protein sequence consists of 20 different types of residues and is encoded using an embedding layer. On the other hand, the protein structure is represented as a sequence of angles, which is encoded using a linear layer. We fuse the sequence and structure embeddings by simply addition:
\begin{equation}
   \label{eq:embedding}
   \begin{cases}
      \rvh^s = \text{Embedding}(\gT(\gS)) \\
      \rvh^x = \text{Linear}(\gT(\gX)) \\
      \rvh^{sx} = \rvh^s + \rvh^x \\
   \end{cases}
\end{equation}
A transformer equipped with rotary position encoding is used to encode token sequence $\rvh^{sx}$ as the continuous embedding $\rvh$:
\begin{equation}
   \label{eq:bert}
   \rvh = \text{Transformer}_{enc}(\rvh^{sx}) 
\end{equation}

\paragraph{Strategy 1: Vanilla Spherical Quantizer} The quantizer maps the continuous embedding $\vh_i$ into the nearest discrete latent codes $\ve_{\hat{i}}$. The quantizer first project high-dimensional $\vh_i \in \sR^{d}$ into a low-dimensional space $\hat{\vh}_i \in \sR^{d'}$ by a linear layer, then normalize the $\hat{\vh}_i$ to the unit hypersphere $\hat{\vh}_i \leftarrow \hat{\vh}_i / ||\hat{\vh}_i||_2$. The the nearest code $\ve_{\hat{i}} \in \sS^{d'}$ in the codebook $\gE$ is found by the following equation:
\begin{equation}
   \label{eq:quantizer}
   \vz_i = \ve_{\hat{i}}, \quad \text{where} \quad \hat{i} = \argmin_{j} ||\vh_i - \vz_j||_2
\end{equation}
We also introduce an EMA updating mechanism to moving $\ve_{i}$ to the center of its nearest top-$k$ embeddings $\{\vh_j: 1 \leq j \leq k\}$, if $\ve_{i}$ is out of the data distribution:
\begin{equation}
   \label{eq:quantizer_ema}
   \ve_{i}  \leftarrow \tau \ve_{i} +  (1-\tau)(\vh_i - \ve_{i}), \quad \text{where} \quad \tau = \begin{cases}
      0.95 & \text{if} \quad || \frac{1}{k}\sum_{j=1}^k \vh_j - \ve_{i}||_2 > \epsilon \\
      0.0 & \text{otherwise} \\
   \end{cases}
\end{equation}
where we choose $k=5$ and $\epsilon=0.1$ in our experiments. We adopt the vq loss to force $\vh_i$ to be close to $\ve_i$: $\mathcal{L}_{vq}=\sum_{i=1}^n \left[ ||\text{sg}(\vh_i) - \ve_{\hat{i}}||_2^2 + \beta ||\vh_i - \text{sg}(\ve_{\hat{i}})||_2^2 \right]$

\paragraph{Strategy 2: Soft Spherical Quantizer} We also propose a soft quantizer to relax the quantization as a continual optimization problem. Given the codebook $\gE$ containing $m$ vectors, the key idea is to query $\gE$ via attention:
\begin{align*}
   a_{ij} &= \frac{\exp{(\hat{\vh}_i^T \ve_j/T)}}{\sum_{j=1}^m \exp{(\hat{\vh}_i^T \ve_j/T)}}\\
   \vz_i &= \sum_{j=1}^m a_{ij} \ve_j
\end{align*}
the temperature $T$ controls the softness of the attention map, i.e., when $T \rightarrow 0$, the attention map becomes a one-hot vector, which is equivalent to the vanilla quantizer. In our experiments, we initially set the temperature parameter $T$ to 1.0 and progressively decrease it to $1e^{-5}$ during the training process. In contrast to the vanilla quantizer, the soft quantizer eliminates the vq loss ($L_{vq}$) and avoids the requirement of gradient estimation on the hidden vector $\hat{\vh}_i^T$. During the inference phase, we randomly sample the discrete latent codes $\rvz$ from the distribution $\texttt{Multinomial}([a_{i,0}, a_{i,1}, \cdots, a_{i,m}])$.

\paragraph{Protein Decoder} The transformer decoder takes the discrete latent codes $\rvz$ as input and reconstructs sequence and structure by separate predictive heads:
\begin{equation}
   \label{eq:decoder}
   \begin{cases}
      \rvh^{dec} = \text{Transformer}_{dec}(\rvz)\\ 
      \hat{\gS} = \text{Linear}_{s}(\rvh^{dec}) \\
      \hat{\gX} = \text{Linear}_{x}(\rvh^{dec}) \\
   \end{cases}
\end{equation}
\paragraph{Overall Loss} The learning objective is to minimize the reconstruction loss between the corrupted proteins and ground-truth ones, including both sequence and structure information. To handle the periodicity of the angles ($\alpha_i, \beta_i$), we minimize their trigonometric values, i.e., $l_{\alpha_i} = || \sin{\alpha_i} - \sin{\hat{\alpha}_i} ||_2^2 +|| \cos{\alpha_i} - \cos{\hat{\alpha}_i} ||_2^2$ for $\alpha_i$.
\begin{align*}
   \label{eq:reconstruction_loss}
   \log p(\gS, \gX | \rvz) &= \log p(\gS | \rvz) + \log p(\gX | \rvz) \\
        &=  -\sum_{i=1}^{n} s_i \log(\hat{s}_i) + \sum_{i=1}^{n}\left(|| r_i - \hat{r}_i ||_2^2 + l_{\alpha_i} + l_{\beta_i} \right)\\
\end{align*}

\section{Experiments}
We conduct experiments to answer the following questions:
\begin{itemize}[leftmargin=5.5mm]
   \item \textbf{Reconstruction (Q1):} How to achieve better reconstruction performance?
   \item \textbf{Predictive Task (Q2):} Does the protein language help to learn expressive protein representations?
   \item \textbf{Generative Task (Q3):} Could the protein language be used for protein generation? (On Going)
\end{itemize}

\subsection{Datasets}
\paragraph{cAF2DB} We use clsutered AlphaFold Uniprot v3 database for pretraining. The original AF2DB contains 214,684,311 structures, which is too large to be used for pretraining. To reduce the computational cost, we adopt the clustered AF2DB (cAF2DB) \citep{barrio2023clustering} that contains 2.27M non-singleton structural clusters. For each cluster, the protein structure with the highest pLDDT score was selected as the representative. We further filter the representative structures by removing ones with lengths less than 30 and pLDDT lower than 70. Finally, we get 1,323,729 proteins for pretraining.

\paragraph{Property Dataset} Following \citep{hermosilla2020intrinsic, hermosilla2022contrastive, zhang2022protein, fan2022continuous}, we assess the efficacy of the developed protein language in four predictive tasks: protein fold classification, enzyme reaction classification, gene ontology (GO) term prediction, and enzyme commission (EC) number prediction. The protein fold classification task encompasses three evaluation scenarios: fold, superfamily, and family. Additionally, the GO term prediction task comprises the prediction of biological process (BP), molecular function (MF), and cellular component (CC) ontology terms.

\paragraph{CATH4.3} We use CATH4.3 for the task of backbone inpainting. We use the CAT code to randomly partition proteins at the ratio of 95:2:3 to construct the train (30290 samples), validation (638 samples) and test sets (957 samples).

\subsection{Pretraining for Reconstruction (Q1)}
\paragraph{Objective \& Setting} The VQProteinformer is pretrained to recover the corrupted protein sequence and structure. We mask 15\% residues in both structure and sequence, as well as add 0.1 Gaussian perturbations to the  Cartesian coordinates of the structure. The sequence reconstruction is straightforward and measured in terms of recovery rate. As to structure reconstruction, we measure from global and local perspectives. The global metric is the TMScore between the recovered and referenced structures. The local metric is the recovering loss per residue, which is defined as $L_{r} = \sum_{i=1}^n | r_i-\hat{r}_i|/n$, $L_{\alpha} = \sum_{i=1}^n | \alpha_i-\hat{\alpha}_i|/n$, and $L_{\beta} = \sum_{i=1}^n | \beta_i-\hat{\beta}_i|/n$. We also report the maximum recovering loss per residue, i.e., $\max L_{r} = \max_{i=1}^n | r_i-\hat{r}_i|$, $\max L_{\alpha} = \max_{i=1}^n | \alpha_i-\hat{\alpha}_i|$, and $\max L_{\beta} = \max_{i=1}^n | \beta_i-\hat{\beta}_i|$.

\paragraph{Pretraining Details} The encoder, decoder and generator are stantard bert models equipped with RoPE (Rotary Position Encoding). We train all these models up to 15 epochs using the OneCycle schedular and AdamW optimizer.

\paragraph{Datasets} We adopt a strict and rigorous protocol: pretraining on cAF2DB and measuring reconstruction performance on the test set of CATH4.3. This choice is motivated by the fact that cAF2DB is generated by AlphaFold, which may introduce unknown biases. By evaluating our model on real proteins from the CATH4.3 dataset, we can gain valuable insights into its generalization ability and assess its performance in a more realistic context.


\begin{table}[h]
   \centering
   \resizebox{1.0 \columnwidth}{!}{
   \begin{tabular}{cccccccccccccccccccc}
   \toprule
   Type-ID & \multicolumn{5}{c}{Config} & \multicolumn{9}{c}{Metrics}     \\ 
   & \#Enc   & \#Dec &\#Codebook   & Noise & Mask   & Rec    & TMScore  & $L_{vq}$ & $L_r$  & $L_{\alpha}$ & $L_{\beta}$  & $\max L_r$  & $\max L_{\alpha}$ & $\max L_{\beta}$   \\ \midrule
   Vanilla-1 &12      & 12   & 1024   & 0.0    & 0.15  & 0.9363 & \textbf{0.4823} & 0.02 & \textbf{0.0960} & \textbf{0.0595} & 0.1041 & \textbf{2.1985} & \textbf{0.5740} & \textbf{0.5384}\\
   Vanilla-2 &6       & 12   & 1024   & 0.0    & 0.15  & \textbf{0.9508} & 0.4240 & 0.02 & 0.0979 & 0.0614 & 0.1107 & 2.1989 & 0.5756 & 0.5508\\
   Vanilla-3 &12      & 6    & 1024   & 0.0    & 0.15  & 0.9453 & 0.1356 & 0.02 & 2.9295 & 2.8682 & \textbf{0.0981} & 5.0615 & 3.0158 & 0.5044\\
   Vanilla-4 &6       & 18   & 1024   & 0.0    & 0.15  & 0.8968 & 0.4427 & 0.02 & 0.0974 & 0.0609 & 0.1205 & 2.2092 & 0.5856 & 0.6482\\ \hline
   Vanilla-5 &12      & 12   & 1024   & 0.0    & 0.15  & 0.9363 & 0.4823 & 0.02 & \textbf{0.0960} & \textbf{0.0595} & 0.1041 & \textbf{2.1985} & \textbf{0.5740} & 0.5384\\
   Vanilla-6 &12      & 12  & 1024    & 0.1    & 0.15  & 0.9416 & \textbf{0.4844} & 0.02 & 0.1236 & 0.0872 & \textbf{0.1015} & 2.2601 & 0.6368 & 0.5286\\ 
   Vanilla-7 &12      & 12  & 1024    & 0.2    & 0.15  & \textbf{0.9443} & 0.4605 & 0.02 & 0.1826 & 0.1464 & 0.1071 & 2.3400 & 0.7249 & \textbf{0.5137}\\  \hline
   Vanilla-8 &12      & 12  & 1024    & 0.1    & 0.00  & \textbf{0.9620} & 0.4521 & 0.02 & \textbf{0.1011} & \textbf{0.0646} & 0.1081 & \textbf{2.2192} & \textbf{0.5965} & 0.5521\\ 
   Vanilla-9 &12      & 12  & 1024    & 0.1    & 0.05  & 0.9565 & 0.4729 & 0.02 & 0.1166 & 0.0801 & 0.1104 & 2.2568 & 0.6344 & 0.6073\\ 
   Vanilla-10 &12      & 12  & 1024    & 0.1    & 0.15  & 0.9416 & \textbf{0.4844} & 0.02 & 0.1236 & 0.0872 & \textbf{0.1015} & 2.2601 & 0.6368 & 0.\textbf{5286}\\ 
   Vanilla-11 &12      & 12  & 1024    & 0.1    & 0.25  & 0.8912 & 0.4212 & 0.02 & 0.1099 & 0.0733 & 0.1195 & 2.2254 & 0.6025 & 0.5577\\ \hline
   Vanilla-12 &12      & 12  & 2048    & 0.1    & 0.15  & 0.9567 & 0.4863 & 0.02 & 0.1190 & 0.0827 & 0.0957 & 2.2421 & 0.6273 & 0.5051\\ \hline
   Soft-1 &12      & 12  & 1024    & 0.1    & 0.15  & 0.5024 & 0.5172 & -- & 0.1567 & 0.1198 & \textbf{0.0829} & 2.4725 & 0.8470 & \textbf{0.3092}\\
   Soft-2 &12      & 12  & 1024    & 0.0    & 0.0  & \textbf{0.9758} & \textbf{0.5743} & -- & \textbf{0.0952} & \textbf{0.0580} & \textbf{0.0881} & \textbf{2.2303} & \textbf{0.5786} & 0.5592\\
   \bottomrule           
   \end{tabular}}
   \caption{Performance of different VQProteinformer models on CATH4.3 test set. We highlight the \textbf{best} results in each group.}
   \label{tab:vq}
\end{table}

\vspace{-3mm}
\paragraph{Results \& Analysis} From the analysis presented in Table.\ref{tab:vq}, we observe that:
\begin{enumerate}
   \item Enhancing the structural similarity (TMScore) proves to be more challenging compared to sequence recovery (Rec).
   \item Both the encoder and decoder components (Vanilla-1 to Vanilla-4) are crucial for protein reconstruction, with the decoder playing a more significant role.
   \item Slightly increasing the scale of structural noise (Vanilla-5 to Vanilla-7) and sequence mask ratio (Vanilla-8 to Vanilla-11) can be advantageous for improving the TMScore for vanilla VQProteinformer.
   \item Introducing a larger vocabulary (Vanilla-8) can lead to a marginal increase in TMScore.
   \item The proposed soft quantizer (Soft-1, Soft-2) achieves significant improvement in TMScore over the vanilla quantizer.
\end{enumerate}

\begin{figure}[h]
   \centering
   \includegraphics[width=4.3in]{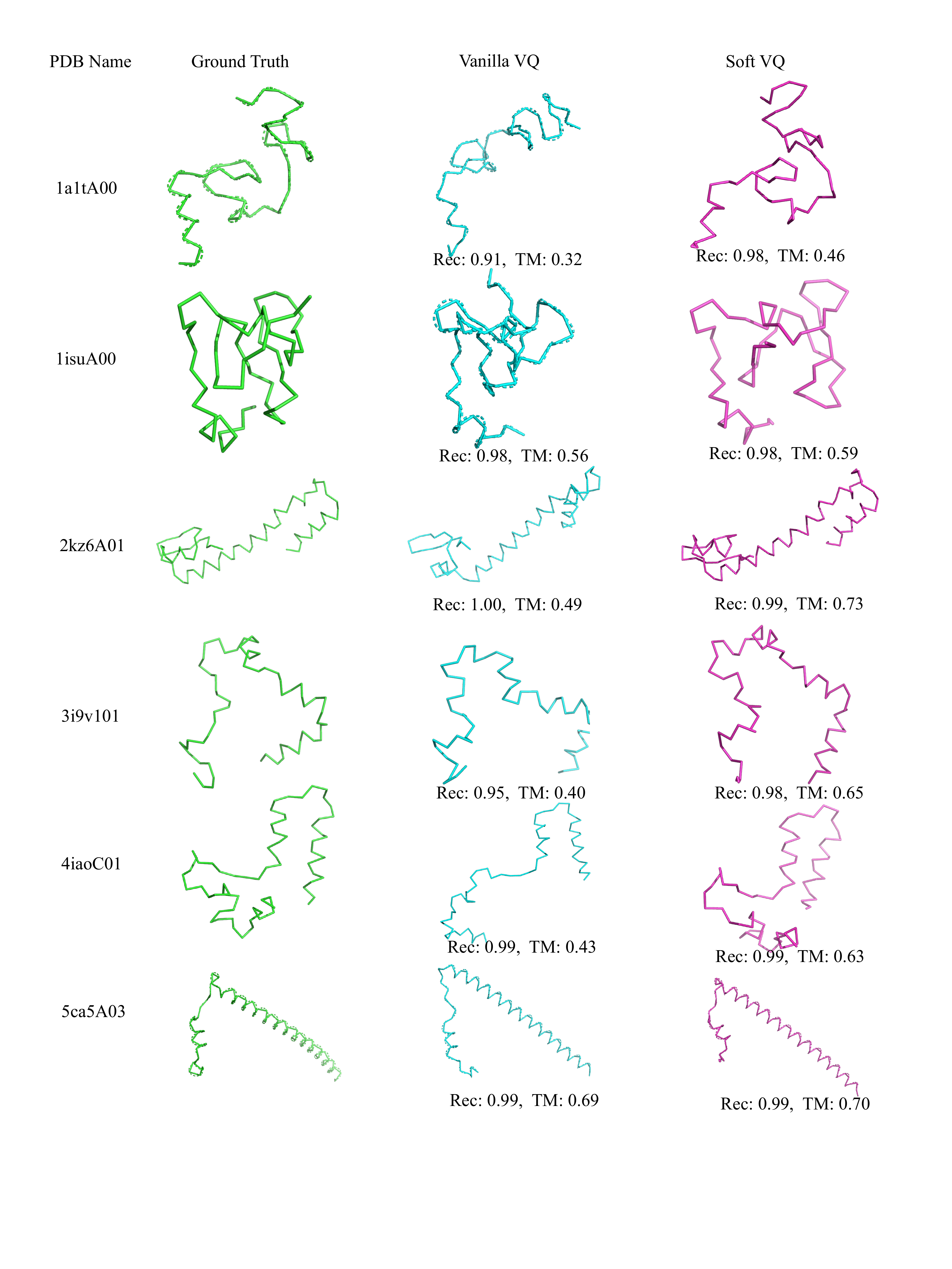}
   \caption{Reconstructed examples. }
   \label{fig:VQ_example}
   \vspace{-5mm}
\end{figure}

\paragraph{Visualization} We compare the reconstructed protein structure with the ground truth in Fig.\ref{fig:VQ_example}. The structures reconstructed by soft VQProteinformer exhibit a higher degree of visual similarity to the reference structures compared to those reconstructed by the vanilla VQProteinformer.

\subsection{Predictive Task (Q2)}
\paragraph{Objective \& Setting} We investigate the efficacy of pretrained embeddings in enhancing downstream predictive tasks. We utilize CDConv as our baseline due to its elegant code and state-of-the-art performance. By substituting the original residue type embedding with pretrained embeddings from GearNet or VQProteinformer, we can assess the influence of pretrained embeddings on the performance. While altering the dimension of the input interface layer, we maintain the model architectures of CDConv consistent with their original paper. In our experimentation, we utilize the latent embedding $\rvh$ from the Vanilla-10 model for the vanilla VQProteinformer. For the soft VQProteinformer, we train a new encoder with the same architectures for representation learning. The encoder is trained with MLM loss to reconstruct the quantized latent codes $\rvz$ of the Soft-2 model. The input sequences are subjected to a 15\% masking ratio and 0.1 Gaussian noise is added to the structures. We evaluate the multiview contrast GearNet in our study.

\paragraph{Results \& Analysis} From Tab.~\ref{tab:predictive}, we observe that:
\begin{enumerate}
   \item Models pretrained on 1M structures do not show significant improvement over the baseline.
   \item The GearNet embeddings could not outperform the baseline in most cases.
   \item The VQProteinformer embeddings slightly outperform the baseline in 6/8 tasks.
   \item Two-stage pretrained model (Soft-2 $\rightarrow$ new MLM encoder) performs worse than one-stage pretrained model (Vanilla-10).
\end{enumerate}

\begin{table}[h]
   \begin{tabular}{ccccccccc} \toprule
           & \multicolumn{3}{c}{Gene Ontology} & \multirow{2}{*}{\begin{tabular}[c]{@{}c@{}}Enzyme \\ Commission\end{tabular}} & \multicolumn{3}{c}{Fold Classification} & \multirow{2}{*}{\begin{tabular}[c]{@{}c@{}}Enzyme \\ Reaction\end{tabular}} \\ 
           & BP         & MF       & CC        &                                                                               & Fold      & Superfamily     & Family    &                                                                             \\ \hline
   Base    & 0.4356     & \textbf{64.97}    & 0.4845    & \textbf{0.8568}                                                                        & 54.01     & 99.30           & 76.43     & 86.94                                                                       \\
   GearNet & 0.4292     & 64.15    & 0.4801    & 0.8455                                                                        & 53.15     & 99.42           & 74.62     & \textbf{88.12}                                                                       \\
   Vanilla VQ      & 0.4349     & 64.1     & \textbf{0.4851}    & 0.8542                                                                        & \textbf{56.01}     & \textbf{99.86}           & \textbf{76.54}     & 86.85               \\
   Soft VQ  & \textbf{0.4380}  & 64.08  & 0.4826 & 0.8520 & 53.88 & 99.27 & 75.06 & 86.99\\ \bottomrule
   \end{tabular}
   \caption{Performance on predictive tasks.}
   \label{tab:predictive}
\end{table}

\subsection{Generative Task (Q3)}
\paragraph{Objective \& Setting} Given the pretrained VQProteinformer, the sequence of latent codes can be interpreted as a form of protein language describing both sequence and structure information. To explore the generative capabilities of this language, we mask a fragment of the protein sequence and autoregressively generate the missing portion. The length of the masked fragment in a protein of length $L$ is uniformly sampled from $U(5, L/3)$. The inpainting models are trained from CATH4.3 dataset, while the protein sentences are generated from pretrained VQProteinformer. The attention mask and loss function are the same as the GLM model \citep{du2021glm}.

\begin{figure}[h]
   \centering
   \includegraphics[width=4.3in]{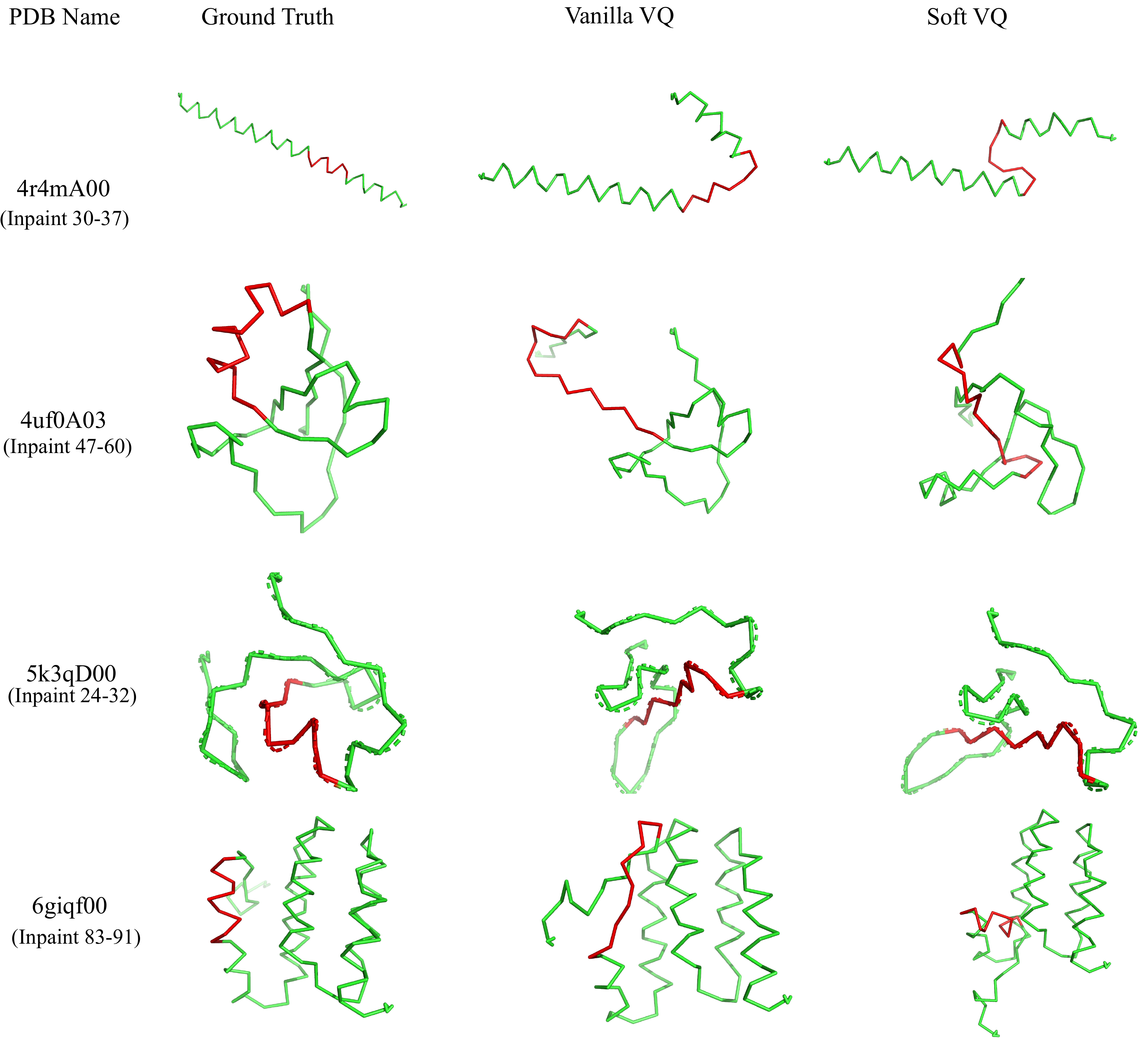}
   \caption{Inpainting examples. }
   \label{fig:Inpainting}
\end{figure}

\paragraph{Results \& Visualization} We currently do not measure the validity of the inpainted proteins, instead, we provide some visual examples in Fig.\ref{fig:Inpainting}.

\section{Conclusion}
In this study, we introduce a fundamental VQProteinformer model that enables the conversion of protein sequences and structures into a discrete "protein language". This novel approach has the potential to revolutionize multimodal protein representation learning and protein design paradigms. While we currently focus on integrating protein sequence and structure modalities, the future holds possibilities for incorporating additional modalities like text and point cloud. The work is currently in progress, and we welcome any suggestions and discussions from the research community.

\bibliography{iclr2022_conference}

\begin{thebibliography}{17}
\providecommand{\natexlab}[1]{#1}
\providecommand{\url}[1]{\texttt{#1}}
\expandafter\ifx\csname urlstyle\endcsname\relax
  \providecommand{\doi}[1]{doi: #1}\else
  \providecommand{\doi}{doi: \begingroup \urlstyle{rm}\Url}\fi

\bibitem[Barrio-Hernandez et~al.(2023)Barrio-Hernandez, Yeo, J{\"a}nes, Wein,
  Varadi, Velankar, Beltrao, and Steinegger]{barrio2023clustering}
Inigo Barrio-Hernandez, Jingi Yeo, J{\"u}rgen J{\"a}nes, Tanita Wein, Mihaly
  Varadi, Sameer Velankar, Pedro Beltrao, and Martin Steinegger.
\newblock Clustering predicted structures at the scale of the known protein
  universe.
\newblock \emph{bioRxiv}, pp.\  2023--03, 2023.

\bibitem[Du et~al.(2021)Du, Qian, Liu, Ding, Qiu, Yang, and Tang]{du2021glm}
Zhengxiao Du, Yujie Qian, Xiao Liu, Ming Ding, Jiezhong Qiu, Zhilin Yang, and
  Jie Tang.
\newblock Glm: General language model pretraining with autoregressive blank
  infilling.
\newblock \emph{arXiv preprint arXiv:2103.10360}, 2021.

\bibitem[Esser et~al.(2021)Esser, Rombach, and Ommer]{esser2021taming}
Patrick Esser, Robin Rombach, and Bjorn Ommer.
\newblock Taming transformers for high-resolution image synthesis.
\newblock In \emph{Proceedings of the IEEE/CVF conference on computer vision
  and pattern recognition}, pp.\  12873--12883, 2021.

\bibitem[Fan et~al.(2022)Fan, Wang, Yang, and Kankanhalli]{fan2022continuous}
Hehe Fan, Zhangyang Wang, Yi~Yang, and Mohan Kankanhalli.
\newblock Continuous-discrete convolution for geometry-sequence modeling in
  proteins.
\newblock In \emph{The Eleventh International Conference on Learning
  Representations}, 2022.

\bibitem[Gao et~al.(2022)Gao, Tan, Li, et~al.]{gao2022alphadesign}
Zhangyang Gao, Cheng Tan, Stan Li, et~al.
\newblock Alphadesign: A graph protein design method and benchmark on
  alphafolddb.
\newblock \emph{arXiv preprint arXiv:2202.01079}, 2022.

\bibitem[Gao et~al.(2023{\natexlab{a}})Gao, Tan, and Li]{gao2023diffsds}
Zhangyang Gao, Cheng Tan, and Stan~Z Li.
\newblock Diffsds: A language diffusion model for protein backbone inpainting
  under geometric conditions and constraints.
\newblock \emph{arXiv preprint arXiv:2301.09642}, 2023{\natexlab{a}}.

\bibitem[Gao et~al.(2023{\natexlab{b}})Gao, Tan, and Li]{gao2023pifold}
Zhangyang Gao, Cheng Tan, and Stan~Z. Li.
\newblock Pifold: Toward effective and efficient protein inverse folding.
\newblock In \emph{International Conference on Learning Representations},
  2023{\natexlab{b}}.
\newblock URL \url{https://openreview.net/forum?id=oMsN9TYwJ0j}.

\bibitem[Hermosilla \& Ropinski(2022)Hermosilla and
  Ropinski]{hermosilla2022contrastive}
Pedro Hermosilla and Timo Ropinski.
\newblock Contrastive representation learning for 3d protein structures.
\newblock \emph{arXiv preprint arXiv:2205.15675}, 2022.

\bibitem[Hermosilla et~al.(2020)Hermosilla, Sch{\"a}fer, Lang, Fackelmann,
  V{\'a}zquez, Kozl{\'\i}kov{\'a}, Krone, Ritschel, and
  Ropinski]{hermosilla2020intrinsic}
Pedro Hermosilla, Marco Sch{\"a}fer, Mat{\v{e}}j Lang, Gloria Fackelmann,
  Pere~Pau V{\'a}zquez, Barbora Kozl{\'\i}kov{\'a}, Michael Krone, Tobias
  Ritschel, and Timo Ropinski.
\newblock Intrinsic-extrinsic convolution and pooling for learning on 3d
  protein structures.
\newblock \emph{arXiv preprint arXiv:2007.06252}, 2020.

\bibitem[Hsu et~al.(2022)Hsu, Verkuil, Liu, Lin, Hie, Sercu, Lerer, and
  Rives]{hsu2022learning}
Chloe Hsu, Robert Verkuil, Jason Liu, Zeming Lin, Brian Hie, Tom Sercu, Adam
  Lerer, and Alexander Rives.
\newblock Learning inverse folding from millions of predicted structures.
\newblock \emph{bioRxiv}, 2022.

\bibitem[Ingraham et~al.(2019)Ingraham, Garg, Barzilay, and
  Jaakkola]{ingraham2019generative}
John Ingraham, Vikas~K Garg, Regina Barzilay, and Tommi Jaakkola.
\newblock Generative models for graph-based protein design.
\newblock 2019.

\bibitem[Jing et~al.(2020)Jing, Eismann, Suriana, Townshend, and
  Dror]{jing2020learning}
Bowen Jing, Stephan Eismann, Patricia Suriana, Raphael~JL Townshend, and Ron
  Dror.
\newblock Learning from protein structure with geometric vector perceptrons.
\newblock \emph{arXiv preprint arXiv:2009.01411}, 2020.

\bibitem[Tan et~al.(2022)Tan, Gao, Xia, and Li]{tan2022generative}
Cheng Tan, Zhangyang Gao, Jun Xia, and Stan~Z Li.
\newblock Generative de novo protein design with global context.
\newblock \emph{arXiv preprint arXiv:2204.10673}, 2022.

\bibitem[Umerenkov et~al.(2022)Umerenkov, Shashkova, Strashnov, Nikolaev,
  Sindeeva, Ivanisenko, and Kardymon]{umerenkov2022prostata}
Dmitriy Umerenkov, Tatiana~I Shashkova, Pavel~V Strashnov, Fedor Nikolaev,
  Maria Sindeeva, Nikita~V Ivanisenko, and Olga~L Kardymon.
\newblock Prostata: Protein stability assessment using transformers.
\newblock \emph{bioRxiv}, pp.\  2022--12, 2022.

\bibitem[Wu et~al.(2022)Wu, Yang, van~den Berg, Zou, Lu, and
  Amini]{wu2022protein}
Kevin~Eric Wu, Kevin~K Yang, Rianne van~den Berg, James Zou, Alex~Xijie Lu, and
  Ava~P Amini.
\newblock Protein structure generation via folding diffusion.
\newblock 2022.

\bibitem[Yu et~al.(2021)Yu, Li, Koh, Zhang, Pang, Qin, Ku, Xu, Baldridge, and
  Wu]{yu2021vector}
Jiahui Yu, Xin Li, Jing~Yu Koh, Han Zhang, Ruoming Pang, James Qin, Alexander
  Ku, Yuanzhong Xu, Jason Baldridge, and Yonghui Wu.
\newblock Vector-quantized image modeling with improved vqgan.
\newblock \emph{arXiv preprint arXiv:2110.04627}, 2021.

\bibitem[Zhang et~al.(2022)Zhang, Xu, Jamasb, Chenthamarakshan, Lozano, Das,
  and Tang]{zhang2022protein}
Zuobai Zhang, Minghao Xu, Arian Jamasb, Vijil Chenthamarakshan, Aurelie Lozano,
  Payel Das, and Jian Tang.
\newblock Protein representation learning by geometric structure pretraining.
\newblock \emph{arXiv preprint arXiv:2203.06125}, 2022.

\end{thebibliography}
\bibliographystyle{iclr2022_conference}

\clearpage
\appendix
\section{Appendix}
\begin{figure}[h]
   \centering
   \includegraphics[width=3.5in]{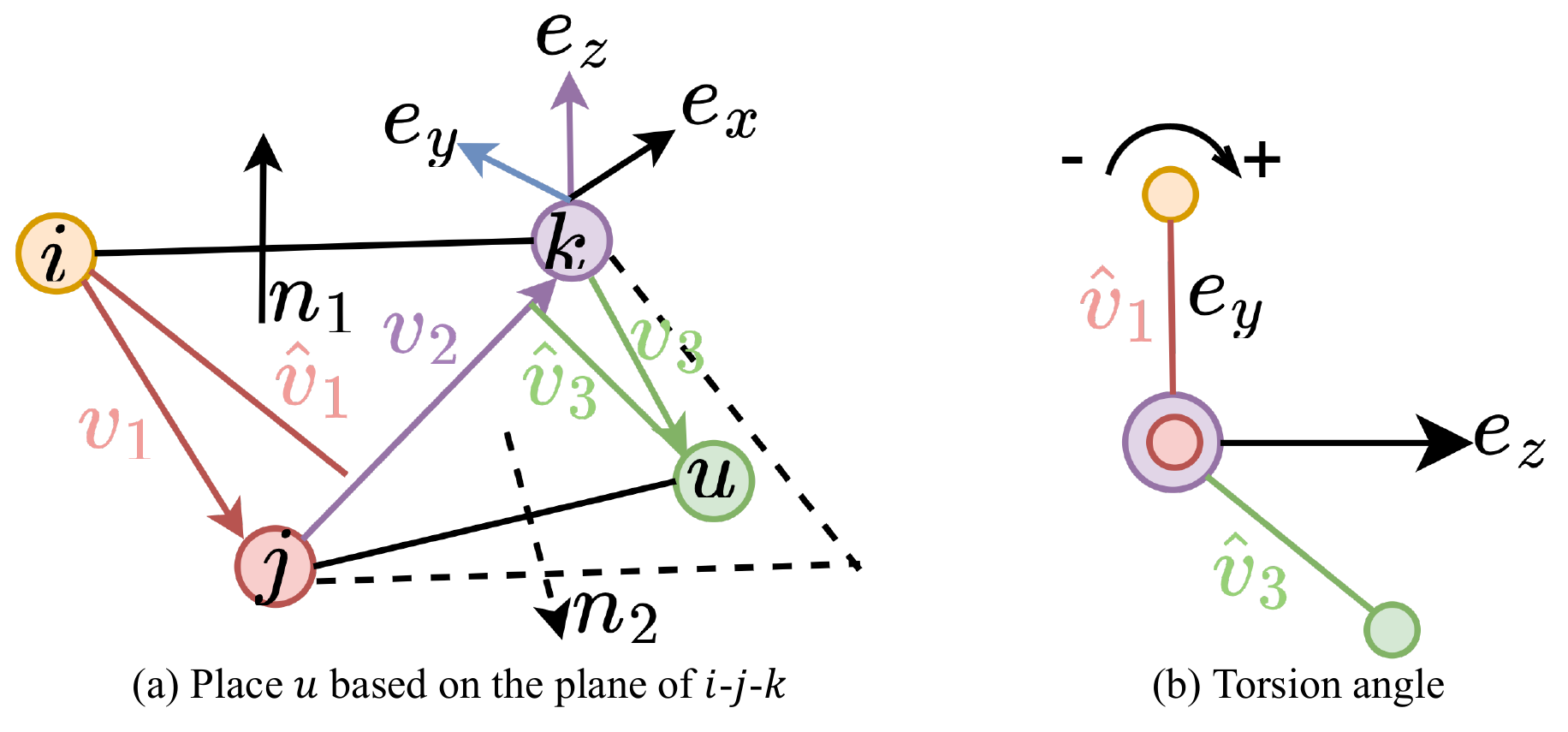}
   \caption{Torsion angle computation. }
   \label{fig:torsion_angle2}
\end{figure}
\paragraph{Structure as angle sequence} The vanilla protein structures are represented as 3d points, where the rotation and translation equivariance should be considered in modeling. To directly apply the transformer-based encoder and decoder, we convert the 3d points into a sequence of angles, which is rotation and translation invariant. As shown in Fig.~\ref{fig:torsion_angle2}, following the backbone direction, given previous three atoms, we can construct a local frame $T_{u} = \{\vx_i, \vx_j, \vx_k\}$. The next atom $\vx_u$ can be represented by the distance $r_u$ and two torsion angles $\alpha_u$ and $\beta_u$, denoted as $\va_u = (r_u, \alpha_u, \beta_u)$. 
\begin{equation}
   \label{eq:torsion_angle}
   \begin{cases}
      \vv_1 = \vx_j - \vx_i \\
      \vv_2 = \vx_k - \vx_j \\
      \vv_3 = \vx_u - \vx_k \\
      \hat{\vv}_1 = -\vv_1 - ((-\vv_1) \cdot \vv_2) \vv_2 \\
      \hat{\vv}_3 = \vv_3 - (\vv_3 \cdot \vv_2) \vv_2 \\
      r_u = ||\vv_3||_2 \\
      \alpha_u = \angle (-\hat{\vv}_2, \hat{\vv}_3) \\
      \beta_u = \angle (\hat{\vv}_1, \hat{\vv}_3) = \angle (\vn_1, \vn_2)\\
   \end{cases}
\end{equation}

To completely represent the structure, we add virtual frames $T_0$ and $T_{n+1}$ at the start and end of the protein structure, where the virtual atoms have fixed $(r, \alpha, \beta)=(1,1,1)$ to its nearest backbone frames. Finally, the protein structure $\gX = \{\vx_i \in \sR^{3}: 1 \leq i \leq n\}$ will be represented as $\gS = \{(r_i, \alpha_i, \beta_i): 0 \leq i \leq n+1\}$.

\end{document}